\documentclass[aps,prl,preprint,groupedaddress]{revtex4-1}

\usepackage{graphicx}

\begin{document}


\title{Two-Dimensional Spectral Interferometry using the Carrier-Envelope Phase}


\author{Christian Ott}
\author{Michael Sch\"onwald}
\author{Philipp Raith}
\author{Andreas Kaldun}
\author{Yizhu Zhang}
\author{Kristina Meyer}
\author{Thomas Pfeifer}
\affiliation{Max-Planck-Institut f\"ur Kernphysik, Saupfercheckweg 1, D-69117 Heidelberg, Germany}


\date{\today}

\begin{abstract}
Two- and multi-dimensional spectroscopy is used in physics and chemistry to obtain structural and dynamical information that would otherwise be invisible by the projection into a one-dimensional data set such as a single emission or absorption spectrum.  Here, we introduce a qualitatively new two-dimensional spectroscopy method by employing the carrier-envelope phase (CEP).  Instead of measuring spectral vs.~spectral information, the combined application of spectral interferometry and CEP control allows the measurement of otherwise inseparable temporal events on an attosecond time scale.  As a specific example, we apply this general method to the case of attosecond pulse train generation, where it allows to separate contributions of three different sub-cycle electron quantum paths within one and the same laser pulse, resulting in a better physical understanding and quantification of the transition region between cutoff and plateau harmonics. The CEP-dependent separation in time between two full-cycle spaced attosecond pulses was determined to modulate by $\left(54\pm16\right)$ attoseconds.
\end{abstract}




\maketitle



Two-dimensional (2D) spectroscopy is a powerful tool to study the energy-level structure, dynamics and couplings among quantum states in complex systems~\cite{Tanimura1993,Asplund2000,Jonas2003}.  The measurements typically involve the acquisition of optical spectra as a function of dynamical (time-delay) coordinates in a multi-pulse nonlinear optical excitation at perturbative intensities.  One of the time-delays is then converted into the spectral domain via Fourier transform and spectral interferometry \cite{Tokunaga1992,Geindre1994}, resulting in two-dimensional spectral~vs.~spectral contour plots that, for instance, allow separation of homogeneous from inhomogeneous spectral linewidths and the measurement of interstate coupling of electronic excitations.  The technique was recently applied to large biomolecules, where complex energy-flow pathways could be visualized on an ultrafast timescale, such as the energy transfer in a photosynthetic light-harvesting complex~\cite{Brixner2005,Engel2007}.  The first crucial step in spectral interferometry (employed in heterodyne 2D spectroscopy) is typically the Fourier transformation of a spectral interferogram into a 'quasi-time' domain to separate and retrieve the phase of the modulating component.  So far, however, 2D spectroscopy methods have not been conceived to separate and analyze electronic events on attosecond time scales in strong few-cycle laser fields.

In such intense few-cycle laser fields, by contrast, carrier-envelope-phase (CEP) control has been shown to be an important parameter, as it allows to steer electron wavepackets on their natural attosecond timescale~\cite{BALTUVSKA2003} due to the precise temporal definition of an electromagnetic field evolution.  One problem in such experiments is that typically an electron wavefunction follows several quantum trajectories or transition pathways (e.g. ionizing at several peaks of the absolute electric field) and the experimentally observed response is a coherent sum over all these trajectories.  Typically, electronic events separated by a half- or full-optical cycle interfere with each other and prevent an isolated observation of distinct trajectories or their quantum phases.

Here, we present a new kind of 2D spectroscopy to quantitatively separate such intra-pulse strong-field processes on an attosecond time scale by combining the method of CEP control with spectral interferometry.  This combination of techniques allows direct \emph{in-situ} and quantitative temporal access to near-periodic strong-field electron dynamics.   The resulting CEP-spectral interferometry (CEPSI) data provides additional two-dimensional \emph{temporal} information in a manner similar (though qualitatively different) to the way 2D spectroscopy allows the extraction of otherwise inseparable \emph{spectral} features or energy-transfer pathways~\cite{Brixner2005,Engel2007}.  Here we will apply the CEPSI method to the example where the relative phase contributions of three attosecond electron trajectories (leading to an attosecond pulse train, APT) in a few-cycle pulse can be qualitatively and quantitatively disentangled in their relative timing and phase.  However, the method itself can be considered to be general for the study of CEP-dependent interference phenomena and should thus find widespread use in strong-field science and physical-chemistry spectroscopy applications, especially when combined with an additional direct dynamical parameter such as a variable time delay of e.g.~a pump and a probe pulse.

APTs were previously reconstructed using cross-correlation techniques between the attosecond pulses and a moderately intense near-infrared (NIR) fundamental laser field ~\cite{Paul2001,KIENBERGER2004} and theoretically developed~\cite{Muller2002,Kitzler2002,Varju2005,Mairesse2005A}, or detected via direct optical interference of two seperately generated APTs~\cite{Mairesse2005}.  An intrinsic property of few-cycle pulse generated APTs is the transistion from the adiabatic to the non-adiabatic regime~\cite{Salieres1998}, where no longer all pulses in the train can be approximated to be equal in shape.  This regime has previously been studied by using the non-adiabatic saddle-point approximation~\cite{Sansone2004}. More recently, a qualitative interpretation of spectral signatures due to interference of only a few attosecond pulses within one single few-cycle laser pulses was discussed~\cite{Pfeifer2007,Mansten2009}, however without direct access to the temporal phases and relative timings which define the created APTs.  Here, we illustrate the power of the introduced general CEPSI method by separating the phase contributions of individual pulses in an APT, shedding light on the underlying electron trajectory evolution.

In the following we first employ a semi-classical model to describe the CEPSI method for the example of few-pulse APT generation, placing emphasis on the analysis of a non-equally spaced triple pulse.  Subsequently we will use this model to analyze and interpret CEP-dependent experimental spectra.  Finally, we will discuss the additional physical knowledge gained from our CEPSI method: it allows to connect our results to previously known regimes of attosecond-pulse production, while at the same time provides a better understanding of the spectral transition region between CEP-dependent cutoff harmonics and CEP-independent plateau harmonics.


We start out with a model description of an APT: in the time domain, it can be formally written as a coherent sum of the electric fields of single attosecond pulses
\begin{equation}\label{PulseTrainTime}
A\left(t\right)=\sum_n A_n\left(t-t_n\right)\exp\left[i\Phi_n\left(t-t_n\right)\right],
\end{equation}
where $A_n\left(t\right)$ and $\Phi_n\left(t\right)$ are the amplitude and phase of each pulse, and $t_n$ denotes their temporal position. In the spectral domain, interference fringes will enter the intensity distribution whenever the slow varying spectral amplitude functions $\tilde{A}_n \left(\omega\right)$ of different attosecond pulses share a common frequency range. Their intensity modulation can be described as $S(\omega)\propto$
\begin{equation}\label{SpectralInterference}
\sum\limits_{m>n} \tilde{A}_n (\omega)\tilde{A}_m (\omega)
\cos\left[\tilde\Phi_n(\omega)-\tilde\Phi_m(\omega)+\omega\tau_{nm}\right],
\end{equation}
where $\tilde\Phi_n\left(\omega\right)$ is the frequency-dependent spectral phase of the $n$-th pulse and $\tau_{nm}=t_m-t_n$ denotes their temporal difference. A Fourier transform of this resulting spectral intensity modulation in Eq.~(\ref{SpectralInterference}) thus gives access to the spectral phase difference $\tilde\Phi_n\left(\omega\right)-\tilde\Phi_m\left(\omega\right)$. Problems with this method arise whenever multiple contributions are overlapping, as it is the case for an APT with multiple half-cycle spaced ($\tau_{nm}=T/2$) pulses. In this case, the contributions cannot be separated.

In the following, we focus on the specific case of a train of three attosecond pulses generated by (HHG) with a few-cycle strong-field driver pulse to illustrate the method.  In that scenario, two temporal contributions exist at roughly the half-cycle spacing.  The key idea of the CEPSI method is to separate the two overlapping temporal contributions by varying the CEP.  As CEP variation in strong-field processes typically affects the phase of different contributions (e.g.~multiphoton transitions~\cite{ABEL2009} or here, quasi-classical electron trajectories) in a different way, the individual components can be separated.

\begin{figure}
\centering
\includegraphics[width=\linewidth]{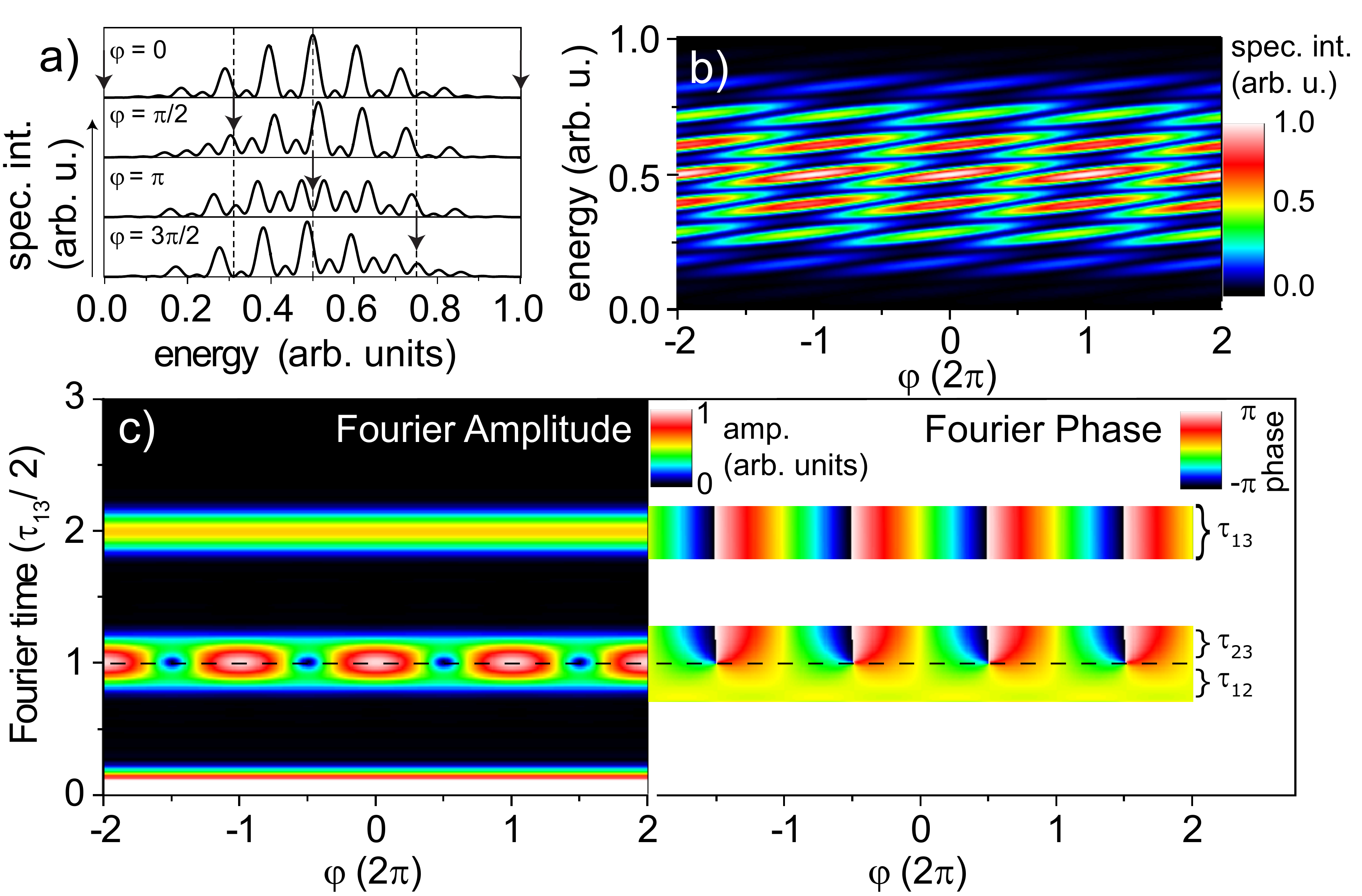}
\caption{\label{FigTripleSlitModel}
(color online) Spectral interference generated by a triple pulse with phases $\Phi_{1,2}=0$ and $\Phi_3=\varphi$ and at non-equidistant temporal spacing $\tau_{12}\approx1.1\tau_{23}$ as a function of the relative phase $\varphi$. a) Spectra for different values of $\varphi$.  The arrows mark the spectral position where the transition from major to minor peaks occurs.  
b) 2d spectroscopic data as a continuous function of energy and relative phase, c) Fourier analysis of b) both in amplitude and phase where the correspondence to the different temporal components $\tau_{12}$, $\tau_{23}$ and $\tau_{13}=\tau_{12}+\tau_{23}$ is indicated.}
\end{figure}

In our illustration model and in analogy to the experimental situation discussed below the temporal spacings $\tau_{12}=t_2-t_1$ and $\tau_{23}=t_3-t_2$ between the three pulses are chosen slightly asymmetrically ($\tau_{23}=\tau_{12}+\Delta$), where $\Delta$ is on the order of the pulse duration, around 10\% of $\tau_{12}$.  The temporal phases of the pulses are chosen to be 0 for the first two pulses and $\varphi$ for the third pulse.  The spectral interference pattern of such a triple pulse is illustrated in Fig.~\ref{FigTripleSlitModel}a) and b).  The spectra consist of alternating major and minor peaks spaced by $2\pi/(\tau_{12}+\tau_{23})$, in analogy to the diffraction pattern of a spatial triple slit. Additionally, for $\Delta\neq 0$, a beating with frequency $2\pi/\Delta$ occurs. At the minima of this beating, the role of the major and minor peaks in the interference pattern swap which is marked by the arrows in Fig.~\ref{FigTripleSlitModel}a). This swapping occurs at different spectral positions, linearly depending on the relative phase $\varphi$ of the third pulse with respect to the other two pulses.  A Fourier analysis of the complete $\varphi$-dependent interference pattern both in amplitude and phase is shown in Fig.~\ref{FigTripleSlitModel}c). Now the previously defined relative pulse characteristics can be separated by observing a $\varphi$-dependent amplitude and phase structure, which would not have been possible for a single value of $\varphi$ or a $\varphi$-averaged spectrum.  The amplitude peak appearing at $\tau_{13}$ exhibits the linear phase difference of $\varphi$ between the two pulses. The amplitude peaks at $\tau_{12}$ and $\tau_{23}$ appear very close to each other within their temporal bandwidth. Due to their different $\varphi$-dependence, a minimum appears in the temporal overlap region at $\varphi=\left(2n+1\right)\pi$, i.e.~when the relative phases between the first two and the latter two pulses are $\pi$ out of phase.


For the experiment, few-cycle ($\sim7\ \textrm{fs}$) laser pulses at 760~nm central wavelength with stable CEP down to $\sim200\ \textrm{mrad}$ are focused, with an $f=500\ \textrm{mm}$ spherical mirror, into a $\sim3$~mm long cell filled with neon gas at $\sim100$~mbar backing pressure. To preferentially phase match short trajectories, the gas cell was placed just after the laser focus~\cite{Antoine1996}. The generated high-harmonic radiation is transmitted through a pair of $200\ \textrm{nm}$ thin zirconium metal foils to remove direct and indirect fundamental stray light.  The remaining coherent soft-x-ray light is spectrally analyzed with a $100\ \textrm{nm}$ period free-standing $\textrm{Si}_3\textrm{N}_4$ transmission grating and detected with a back-illuminated x-ray-CCD camera.

\begin{figure}
\centering
\includegraphics[width=\linewidth]{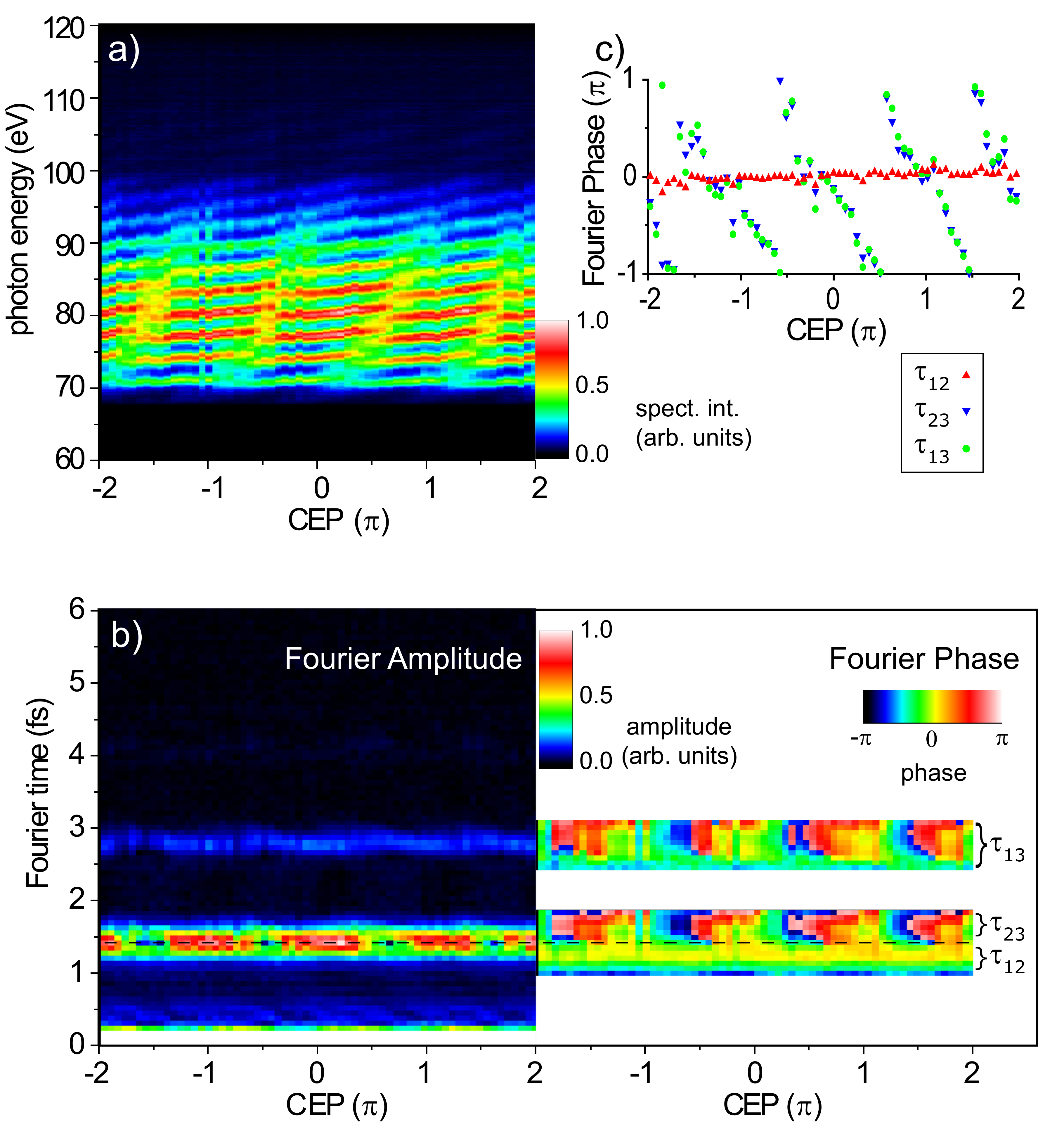}
\caption{\label{FigExperimentSpectra}
(color online) Experimental results: a) Spectral interference pattern recorded as a function of the CEP.  At energies below $95\ \textrm{eV}$ a splitting of the interference structures (harmonics) occurs as discussed in the model of Fig.~\ref{FigTripleSlitModel}. b) Fourier analysis of a):  splitting of the amplitude peak as well as a bimodal phase behaviour is observed as a function of CEP, as in the model presented before.  c) The phase averaged over the marked temporal regions for $\tau_{12}$ (red triangle up), $\tau_{23}$ (blue triangle down) and $\tau_{13}=\tau_{12}+\tau_{23}$ (circle, green) is shown to clarify the qualitatively different CEP dependences.}
\end{figure}

In Fig.~\ref{FigExperimentSpectra}a), spatially integrated experimental spectra are shown as a function of the CEP. 
The spectra exhibit interference structures spaced by $\sim 3\ \textrm{eV}$, the well known $2\omega_0$ odd-harmonic spacing. For the energies above $95\ \textrm{eV}$ the interference fringes shift linearly from one order to the next within one CEP cycle confirming earlier findings in the harmonic cutoff region~\cite{Nisoli2003,Sansone2004}. Interestingly, in the energy region just below the cutoff ($<95$~eV), a switching between different harmonic peaks occurs with a change in CEP.  Repeating the previously illustrated Fourier analysis on these experimental data (Fig.~\ref{FigExperimentSpectra}b) results in a close agreement to the previous asymmetric three-pulse toy model, where the CEP now takes the role of the general phase $\varphi$ as introduced before.  A CEP dependence both for the full-cycle Fourier peak $\left(\tau_{13}\right)$ as well as for the upper temporal part $\left(\tau_{23}\right)$ of the half-cycle Fourier peak appears, whereas the lower temporal part $\left(\tau_{12}\right)$ of this peak shows a constant phase as a function of the CEP. The local minima in the half-cylce Fourier peak in Fig.~\ref{FigExperimentSpectra}b) occur periodically at CEPs of 0.5 modulo $\pi$.  An integration over the phase for the different Fourier components is shown in Fig.~\ref{FigExperimentSpectra}c) to extract the different phase evolutions.


\begin{figure}
\centering
\includegraphics[width=\linewidth]{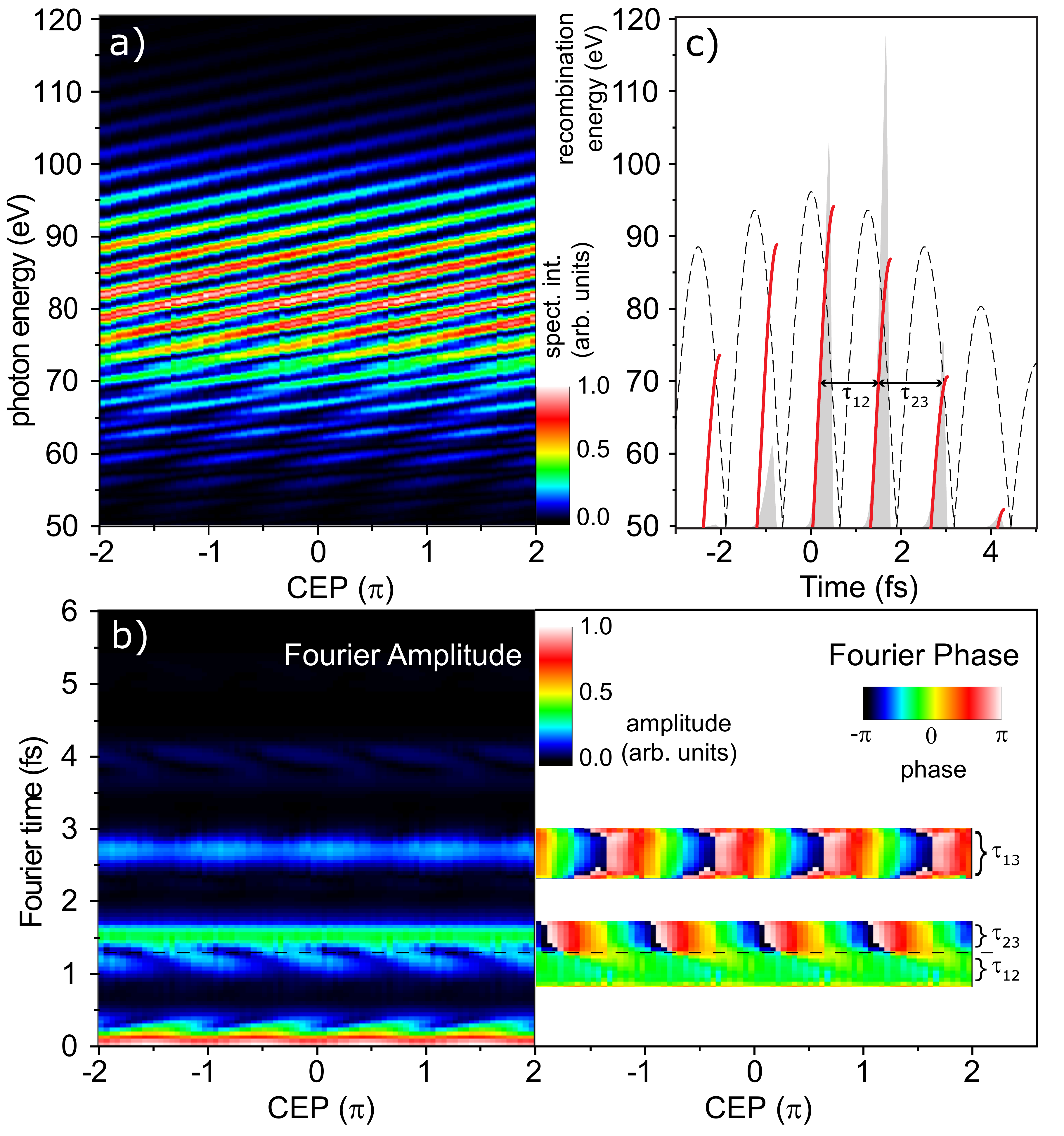}
\caption{\label{FigTrajectories}
(color online) Strong-field approximation simulation results: a) Spectral interference pattern as a function of the CEP.  Below 75~eV, a splitting of harmonic positions at certain CEPs occurs.  b) Fourier analysis of a) both in amplitude and phase, where only the region below 75~eV was considered.  Again, the qualitatively different phase contributions to $\tau_{12}$ and $\tau_{23}$ can be seen, accompanied by a minimum in the amplitude peak.  c) Classically calculated electron recombination energies for $\varphi_{\mathrm{CEP}}=0$ for the short trajectory (red solid line). The absolute value of the IR electric field $\left|E\left(t\right)\right|$ (dashed line) and the attosecond pulse envelopes (gray shaded area) are shown. The different temporal spacings $\tau_{12}$ and $\tau_{23}$ are indicated for the three most intense attosecond pulses in the energy region, where the splitting effect occurs.}
\end{figure}

To better understand the quantum-dynamical mechanism behind the formation of this asymmetric triple slit in the time domain, we carried out a quasi-classical strong-field approximation simulation of HHG based on the three-step model~\cite{Corkum1993,Lewenstein1994}.  Only the short trajectories were considered.  A $7\ \textrm{fs}$ FWHM Gaussian laser pulse centered around $760\ \textrm{nm}$ with peak electric field strength $0.11$~a.u.~was used in the simulations.
The simulated harmonic spectra are shown in Fig.~\ref{FigTrajectories}a).  At lower energies around 65~eV, the characteristic switching of the harmonic peaks becomes apparent, in agreement both with the experiment as well as with the presented initial toy model of non-equidistantly spaced triple pulses.  At this stage it should be pointed out that propagation effects are not taken into account, which may be the reason behind some slight quantitative disagreement between the experiment and the single-atom simulations.

Knowledge and analysis of the quasi-classical electron trajectories and their recollision energetics allows to trace the origin of the observed triple-pulse behavior, as shown in Fig.~\ref{FigTrajectories}c): The three most intense pulses exhibit a different temporal spacing at a given photon energy, indicated by the horizontal arrows $\tau_{12}$ and $\tau_{23}$.  For the two most intense pulses, a photon energy of 65~eV corresponds to their half-cycle-plateau region (strong frequency upchirp of the individual attosecond pulses), whereas  the third pulse contributes to this energy out of its half-cycle cutoff region (no frequency chirp near cutoff). The asymmetric temporal spacing of the three pulses thus clearly originates from the nonlinear chirp of the attosecond pulses that arises when the short trajectory of one pulse smoothly transits from its plateau into its cutoff region. It is well known \cite{BALCOU1999,Sansone2004} that the cutoff-harmonic phase depends roughly linearly on intensity while the plateau harmonics for the short trajectory are approximately independent of the intensity.  By varying the CEP, we move individual attosecond pulse contributions up and down the envelope of the driver pulse, thus changing the intensity with which they are produced on a sub-cycle (attosecond) time scale. CEPSI thus allows to separate these different phase contributions caused by individual trajectories on a sub-cycle basis, surpassing previous approaches that used direct intensity variation of the entire pulse to extract intensity-dependent phases averaged over all half-cycles of the entire driver pulse.  Here, the sub-cycle resolution allowed us to disentangle the trajectories in the transition region from the cutoff to the plateau regime \emph{within} the driver pulse. 

\begin{figure}
\centering
\includegraphics[width=\linewidth]{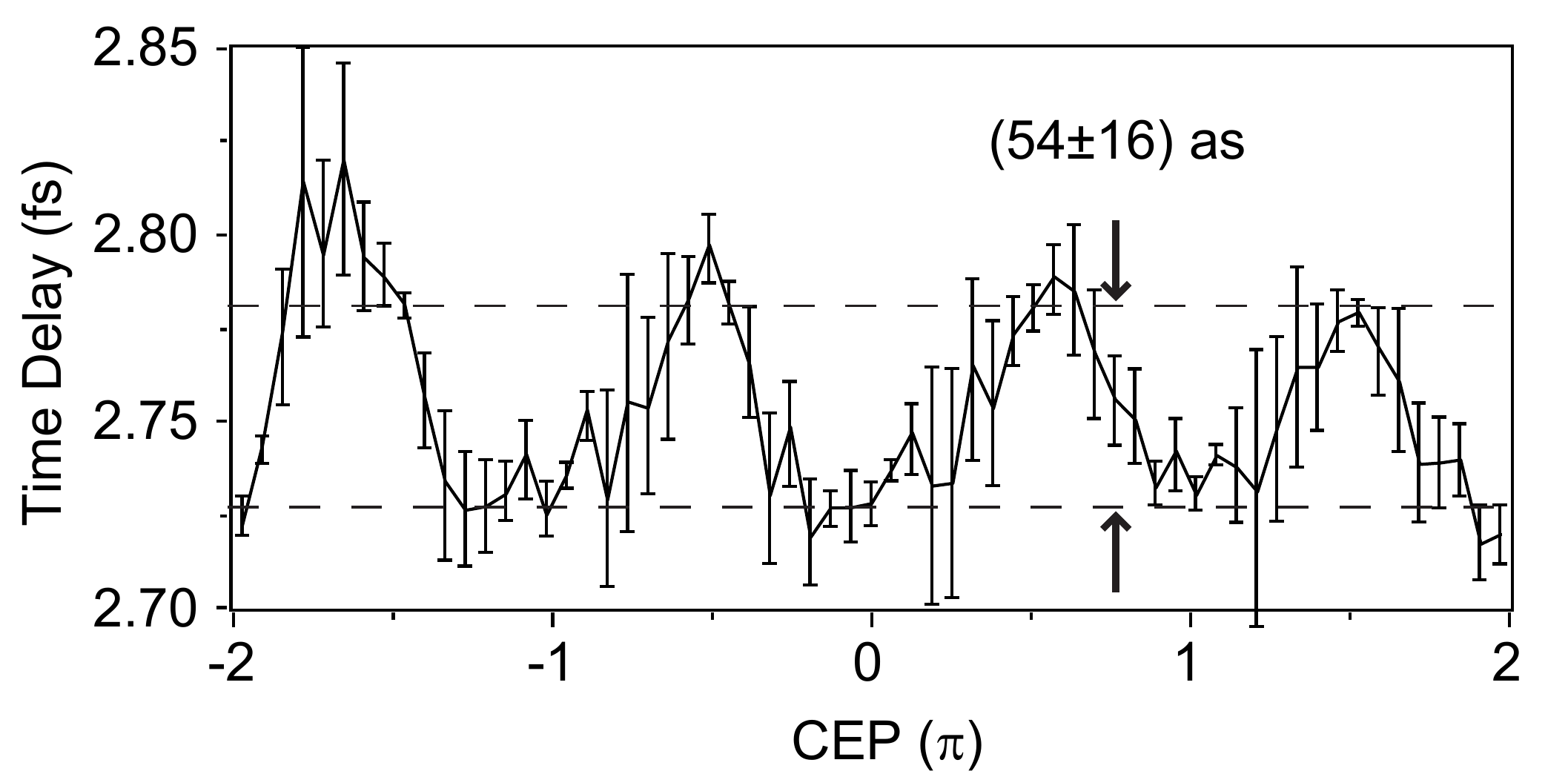}
\caption{\label{FigGroupDelay}
Mean time delay $\tau_{13}=\tau_{12}+\tau_{23}$ of the two outermost pulses in the attosecond-pulse triplet as a function of CEP, extracted from the experimental results in Fig.~\ref{FigExperimentSpectra}. The error bars denote the standard deviation of the Gaussian fits to the $\pi$-periodic data sets.  A fit to the systematic variation in time delay results in  $\left(54\pm16\right)\textrm{attoseconds}$ (horizontal dashed lines) as the third pulse moves into and out of the cutoff region in the spectral range considered here (70-90~eV).}
\end{figure}

Finally, after CEPSI allowed extracting the qualitative physical mechanism, we now further use it to quantify the CEP-dependent relative group delay between two full-cycle spaced attosecond pulses in the experiment.  The full-cycle Fourier peak corresponding to $\tau_{13}=\tau_{12}+\tau_{23}$ in Fig.~\ref{FigExperimentSpectra}b) is fitted with a Gaussian function to determine its central position at each value of the CEP. The result is shown in Fig.~\ref{FigGroupDelay}, where a periodic modulation in time delay of $\left(54\pm16\right)\textrm{attoseconds}$ can be observed. The minima appear at CEP values where all three pulses are approximately equal in phase, whereas the maxima correspond to CEP values where phase-jumps occur across the half-cycle Fourier peak (compare to Fig.~\ref{FigExperimentSpectra}).  An explanation for this modulation is as follows: For minimal group delays, the corresponding CEP values belong to situations where all three pulses contribute with their plateaus. In contrast, when the group delay is at its maximum, the corresponding attosecond pulse configurations consist of only two pulses within their plateau region, while the third pulse on the trailing edge contributes with its own cutoff region (as indicated in Fig.~\ref{FigTrajectories}c)), resulting in a larger spacing $\tau_{13}$ between pulses 2 and 3.

In conclusion we introduced the combination of CEP sweeping and spectral interferometry (CEPSI) to create a novel sub-cycle multidimensional spectroscopy method.  It allows to separate otherwise overlapping quantum paths within the same strong-field driver pulse.  Application of this method in HHG allows access to the transition region between the cutoff and plateau part in the spectrum of APTs by qualitatively and quantitatively separating contributions of individual sub-cycle electron trajectories.  An asymmetric arrangement of three attosecond pulses could thus be identified and analyzed with respect to its CEP dependence and the variation of the relative timing of the two outermost pulses in the pulse triplet could be measured with sub-20 attosecond precision.

While applied here to the special case of HHG as an illustration example, the method is general and can enhance physical insight on other strong field processes that are known to depend on the CEP within a single pulse, such as (above-threshold) ionization in the multiphoton or tunneling regime, recollision imaging, and electron localization in molecules.  In all these processes several different quantum (transition, transport) pathways exist, with their phase depending sensitively on the CEP.  In analogy to the high-harmonic case considered here, CEPSI can separate the contributing quantum paths by their phase dependence and thus extract novel physical mechanisms as well as quantify already existing ones at high temporal or spectral resolution.

Financial support from the Max-Planck Research Group program is gratefully acknowledged.


\begin{thebibliography}{25}
\expandafter\ifx\csname natexlab\endcsname\relax\def\natexlab#1{#1}\fi
\expandafter\ifx\csname bibnamefont\endcsname\relax
  \def\bibnamefont#1{#1}\fi
\expandafter\ifx\csname bibfnamefont\endcsname\relax
  \def\bibfnamefont#1{#1}\fi
\expandafter\ifx\csname citenamefont\endcsname\relax
  \def\citenamefont#1{#1}\fi
\expandafter\ifx\csname url\endcsname\relax
  \def\url#1{\texttt{#1}}\fi
\expandafter\ifx\csname urlprefix\endcsname\relax\def\urlprefix{URL }\fi
\providecommand{\bibinfo}[2]{#2}
\providecommand{\eprint}[2][]{\url{#2}}

\bibitem[{\citenamefont{Tanimura and Mukamel}(1993)}]{Tanimura1993}
\bibinfo{author}{\bibfnamefont{Y.}~\bibnamefont{Tanimura}} \bibnamefont{and}
  \bibinfo{author}{\bibfnamefont{S.}~\bibnamefont{Mukamel}},
  \bibinfo{journal}{J. Chem. Phys.} \textbf{\bibinfo{volume}{99}},
  \bibinfo{pages}{9496} (\bibinfo{year}{1993}).

\bibitem[{\citenamefont{Asplund et~al.}(2000)\citenamefont{Asplund, Zanni, and
  Hochstrasser}}]{Asplund2000}
\bibinfo{author}{\bibfnamefont{M.~C.} \bibnamefont{Asplund}},
  \bibinfo{author}{\bibfnamefont{M.~T.} \bibnamefont{Zanni}}, \bibnamefont{and}
  \bibinfo{author}{\bibfnamefont{R.~M.} \bibnamefont{Hochstrasser}},
  \bibinfo{journal}{Proc. Natl. Acad. Sci. U. S. A.}
  \textbf{\bibinfo{volume}{97}}, \bibinfo{pages}{8219} (\bibinfo{year}{2000}).

\bibitem[{\citenamefont{Jonas}(2003)}]{Jonas2003}
\bibinfo{author}{\bibfnamefont{D.~M.} \bibnamefont{Jonas}},
  \bibinfo{journal}{Annu. Rev. Phys. Chem.} \textbf{\bibinfo{volume}{54}},
  \bibinfo{pages}{425} (\bibinfo{year}{2003}).

\bibitem[{\citenamefont{Tokunaga et~al.}(1992)\citenamefont{Tokunaga, Terasaki,
  and Kobayashi}}]{Tokunaga1992}
\bibinfo{author}{\bibfnamefont{E.}~\bibnamefont{Tokunaga}},
  \bibinfo{author}{\bibfnamefont{A.}~\bibnamefont{Terasaki}}, \bibnamefont{and}
  \bibinfo{author}{\bibfnamefont{T.}~\bibnamefont{Kobayashi}},
  \bibinfo{journal}{Opt. Lett.} \textbf{\bibinfo{volume}{17}},
  \bibinfo{pages}{1131} (\bibinfo{year}{1992}).

\bibitem[{\citenamefont{Geindre et~al.}(1994)\citenamefont{Geindre, Audebert,
  Rousse, Falli\`{e}s, Gauthier, Mysyrowicz, Santos, Hamoniaux, and
  Antonetti}}]{Geindre1994}
\bibinfo{author}{\bibfnamefont{J.~P.} \bibnamefont{Geindre}},
  \bibinfo{author}{\bibfnamefont{P.}~\bibnamefont{Audebert}},
  \bibinfo{author}{\bibfnamefont{A.}~\bibnamefont{Rousse}},
  \bibinfo{author}{\bibfnamefont{F.}~\bibnamefont{Falli\`{e}s}},
  \bibinfo{author}{\bibfnamefont{J.~C.} \bibnamefont{Gauthier}},
  \bibinfo{author}{\bibfnamefont{A.}~\bibnamefont{Mysyrowicz}},
  \bibinfo{author}{\bibfnamefont{A.~D.} \bibnamefont{Santos}},
  \bibinfo{author}{\bibfnamefont{G.}~\bibnamefont{Hamoniaux}},
  \bibnamefont{and}
  \bibinfo{author}{\bibfnamefont{A.}~\bibnamefont{Antonetti}},
  \bibinfo{journal}{Opt. Lett.} \textbf{\bibinfo{volume}{19}},
  \bibinfo{pages}{1997} (\bibinfo{year}{1994}).

\bibitem[{\citenamefont{Brixner et~al.}(2005)\citenamefont{Brixner, Stenger,
  Vaswani, Cho, Blankenship, and Fleming}}]{Brixner2005}
\bibinfo{author}{\bibfnamefont{T.}~\bibnamefont{Brixner}},
  \bibinfo{author}{\bibfnamefont{J.}~\bibnamefont{Stenger}},
  \bibinfo{author}{\bibfnamefont{H.~M.} \bibnamefont{Vaswani}},
  \bibinfo{author}{\bibfnamefont{M.}~\bibnamefont{Cho}},
  \bibinfo{author}{\bibfnamefont{R.~E.} \bibnamefont{Blankenship}},
  \bibnamefont{and} \bibinfo{author}{\bibfnamefont{G.~R.}
  \bibnamefont{Fleming}}, \bibinfo{journal}{Nature}
  \textbf{\bibinfo{volume}{434}}, \bibinfo{pages}{625} (\bibinfo{year}{2005}).

\bibitem[{\citenamefont{Engel et~al.}(2007)\citenamefont{Engel, Calhoun, Read,
  Ahn, Mancal, Cheng, Blankenship, and Fleming}}]{Engel2007}
\bibinfo{author}{\bibfnamefont{G.~S.} \bibnamefont{Engel}},
  \bibinfo{author}{\bibfnamefont{T.~R.} \bibnamefont{Calhoun}},
  \bibinfo{author}{\bibfnamefont{E.~L.} \bibnamefont{Read}},
  \bibinfo{author}{\bibfnamefont{T.~K.} \bibnamefont{Ahn}},
  \bibinfo{author}{\bibfnamefont{T.}~\bibnamefont{Mancal}},
  \bibinfo{author}{\bibfnamefont{Y.~C.} \bibnamefont{Cheng}},
  \bibinfo{author}{\bibfnamefont{R.~E.} \bibnamefont{Blankenship}},
  \bibnamefont{and} \bibinfo{author}{\bibfnamefont{G.~R.}
  \bibnamefont{Fleming}}, \bibinfo{journal}{Nature}
  \textbf{\bibinfo{volume}{446}}, \bibinfo{pages}{782} (\bibinfo{year}{2007}).

\bibitem[{\citenamefont{Baltu\v{s}ka et~al.}(2003)\citenamefont{Baltu\v{s}ka,
  Udem, Uiberacker, Hentschel, Goulielmakis, Gohle, Holzwarth, Yakovlev,
  Scrinzi, H\"ansch et~al.}}]{BALTUVSKA2003}
\bibinfo{author}{\bibfnamefont{A.}~\bibnamefont{Baltu\v{s}ka}},
  \bibinfo{author}{\bibfnamefont{T.}~\bibnamefont{Udem}},
  \bibinfo{author}{\bibfnamefont{M.}~\bibnamefont{Uiberacker}},
  \bibinfo{author}{\bibfnamefont{M.}~\bibnamefont{Hentschel}},
  \bibinfo{author}{\bibfnamefont{E.}~\bibnamefont{Goulielmakis}},
  \bibinfo{author}{\bibfnamefont{C.}~\bibnamefont{Gohle}},
  \bibinfo{author}{\bibfnamefont{R.}~\bibnamefont{Holzwarth}},
  \bibinfo{author}{\bibfnamefont{V.~S.} \bibnamefont{Yakovlev}},
  \bibinfo{author}{\bibfnamefont{A.}~\bibnamefont{Scrinzi}},
  \bibinfo{author}{\bibfnamefont{T.~W.} \bibnamefont{H\"ansch}},
  \bibnamefont{et~al.}, \bibinfo{journal}{Nature}
  \textbf{\bibinfo{volume}{421}}, \bibinfo{pages}{611} (\bibinfo{year}{2003}).

\bibitem[{\citenamefont{Paul et~al.}(2001)\citenamefont{Paul, Toma, Breger,
  Mullot, Auge, Balcou, Muller, and Agostini}}]{Paul2001}
\bibinfo{author}{\bibfnamefont{P.~M.} \bibnamefont{Paul}},
  \bibinfo{author}{\bibfnamefont{E.~S.} \bibnamefont{Toma}},
  \bibinfo{author}{\bibfnamefont{P.}~\bibnamefont{Breger}},
  \bibinfo{author}{\bibfnamefont{G.}~\bibnamefont{Mullot}},
  \bibinfo{author}{\bibfnamefont{F.}~\bibnamefont{Auge}},
  \bibinfo{author}{\bibfnamefont{P.}~\bibnamefont{Balcou}},
  \bibinfo{author}{\bibfnamefont{H.~G.} \bibnamefont{Muller}},
  \bibnamefont{and} \bibinfo{author}{\bibfnamefont{P.}~\bibnamefont{Agostini}},
  \bibinfo{journal}{Science} \textbf{\bibinfo{volume}{292}},
  \bibinfo{pages}{1689} (\bibinfo{year}{2001}).

\bibitem[{\citenamefont{Kienberger et~al.}(2004)\citenamefont{Kienberger,
  Goulielmakis, Uiberacker, Baltu\v{s}ka, Yakovlev, Bammer, Scrinzi,
  Westerwalbesloh, Kleineberg, Heinzmann et~al.}}]{KIENBERGER2004}
\bibinfo{author}{\bibfnamefont{R.}~\bibnamefont{Kienberger}},
  \bibinfo{author}{\bibfnamefont{E.}~\bibnamefont{Goulielmakis}},
  \bibinfo{author}{\bibfnamefont{M.}~\bibnamefont{Uiberacker}},
  \bibinfo{author}{\bibfnamefont{A.}~\bibnamefont{Baltu\v{s}ka}},
  \bibinfo{author}{\bibfnamefont{V.}~\bibnamefont{Yakovlev}},
  \bibinfo{author}{\bibfnamefont{F.}~\bibnamefont{Bammer}},
  \bibinfo{author}{\bibfnamefont{A.}~\bibnamefont{Scrinzi}},
  \bibinfo{author}{\bibfnamefont{T.}~\bibnamefont{Westerwalbesloh}},
  \bibinfo{author}{\bibfnamefont{U.}~\bibnamefont{Kleineberg}},
  \bibinfo{author}{\bibfnamefont{U.}~\bibnamefont{Heinzmann}},
  \bibnamefont{et~al.}, \bibinfo{journal}{Nature}
  \textbf{\bibinfo{volume}{427}}, \bibinfo{pages}{817} (\bibinfo{year}{2004}).

\bibitem[{\citenamefont{Muller}(2002)}]{Muller2002}
\bibinfo{author}{\bibfnamefont{H.~G.} \bibnamefont{Muller}},
  \bibinfo{journal}{Appl. Phys. B} \textbf{\bibinfo{volume}{74}},
  \bibinfo{pages}{17} (\bibinfo{year}{2002}).

\bibitem[{\citenamefont{Kitzler et~al.}(2002)\citenamefont{Kitzler, Milosevic,
  Scrinzi, Krausz, and Brabec}}]{Kitzler2002}
\bibinfo{author}{\bibfnamefont{M.}~\bibnamefont{Kitzler}},
  \bibinfo{author}{\bibfnamefont{N.}~\bibnamefont{Milosevic}},
  \bibinfo{author}{\bibfnamefont{A.}~\bibnamefont{Scrinzi}},
  \bibinfo{author}{\bibfnamefont{F.}~\bibnamefont{Krausz}}, \bibnamefont{and}
  \bibinfo{author}{\bibfnamefont{T.}~\bibnamefont{Brabec}},
  \bibinfo{journal}{Phys. Rev. Lett.} \textbf{\bibinfo{volume}{88}},
  \bibinfo{pages}{173904} (\bibinfo{year}{2002}).

\bibitem[{\citenamefont{Varj\'u et~al.}(2005)\citenamefont{Varj\'u, Mairesse,
  Agostini, Breger, Carr\'e, Frasinski, Gustafsson, Johnsson, Mauritsson,
  Merdji et~al.}}]{Varju2005}
\bibinfo{author}{\bibfnamefont{K.}~\bibnamefont{Varj\'u}},
  \bibinfo{author}{\bibfnamefont{Y.}~\bibnamefont{Mairesse}},
  \bibinfo{author}{\bibfnamefont{P.}~\bibnamefont{Agostini}},
  \bibinfo{author}{\bibfnamefont{P.}~\bibnamefont{Breger}},
  \bibinfo{author}{\bibfnamefont{B.}~\bibnamefont{Carr\'e}},
  \bibinfo{author}{\bibfnamefont{L.~J.} \bibnamefont{Frasinski}},
  \bibinfo{author}{\bibfnamefont{E.}~\bibnamefont{Gustafsson}},
  \bibinfo{author}{\bibfnamefont{P.}~\bibnamefont{Johnsson}},
  \bibinfo{author}{\bibfnamefont{J.}~\bibnamefont{Mauritsson}},
  \bibinfo{author}{\bibfnamefont{H.}~\bibnamefont{Merdji}},
  \bibnamefont{et~al.}, \bibinfo{journal}{Phys. Rev. Lett.}
  \textbf{\bibinfo{volume}{95}}, \bibinfo{pages}{243901}
  (\bibinfo{year}{2005}).

\bibitem[{\citenamefont{Mairesse and Qu\'er\'e}(2005)}]{Mairesse2005A}
\bibinfo{author}{\bibfnamefont{Y.}~\bibnamefont{Mairesse}} \bibnamefont{and}
  \bibinfo{author}{\bibfnamefont{F.}~\bibnamefont{Qu\'er\'e}},
  \bibinfo{journal}{Phys. Rev. A} \textbf{\bibinfo{volume}{71}},
  \bibinfo{pages}{011401} (\bibinfo{year}{2005}).

\bibitem[{\citenamefont{Mairesse et~al.}(2005)\citenamefont{Mairesse, Gobert,
  Breger, Merdji, Meynadier, Monchicourt, Perdrix, Sali\`eres, and
  Carr\'e}}]{Mairesse2005}
\bibinfo{author}{\bibfnamefont{Y.}~\bibnamefont{Mairesse}},
  \bibinfo{author}{\bibfnamefont{O.}~\bibnamefont{Gobert}},
  \bibinfo{author}{\bibfnamefont{P.}~\bibnamefont{Breger}},
  \bibinfo{author}{\bibfnamefont{H.}~\bibnamefont{Merdji}},
  \bibinfo{author}{\bibfnamefont{P.}~\bibnamefont{Meynadier}},
  \bibinfo{author}{\bibfnamefont{P.}~\bibnamefont{Monchicourt}},
  \bibinfo{author}{\bibfnamefont{M.}~\bibnamefont{Perdrix}},
  \bibinfo{author}{\bibfnamefont{P.}~\bibnamefont{Sali\`eres}},
  \bibnamefont{and} \bibinfo{author}{\bibfnamefont{B.}~\bibnamefont{Carr\'e}},
  \bibinfo{journal}{Phys. Rev. Lett.} \textbf{\bibinfo{volume}{94}},
  \bibinfo{pages}{173903} (\bibinfo{year}{2005}).

\bibitem[{\citenamefont{Sali\`eres et~al.}(1998)\citenamefont{Sali\`eres,
  Antoine, de~Bohan, and Lewenstein}}]{Salieres1998}
\bibinfo{author}{\bibfnamefont{P.}~\bibnamefont{Sali\`eres}},
  \bibinfo{author}{\bibfnamefont{P.}~\bibnamefont{Antoine}},
  \bibinfo{author}{\bibfnamefont{A.}~\bibnamefont{de~Bohan}}, \bibnamefont{and}
  \bibinfo{author}{\bibfnamefont{M.}~\bibnamefont{Lewenstein}},
  \bibinfo{journal}{Phys. Rev. Lett.} \textbf{\bibinfo{volume}{81}},
  \bibinfo{pages}{5544} (\bibinfo{year}{1998}).

\bibitem[{\citenamefont{Sansone et~al.}(2004)\citenamefont{Sansone, Vozzi,
  Stagira, and Nisoli}}]{Sansone2004}
\bibinfo{author}{\bibfnamefont{G.}~\bibnamefont{Sansone}},
  \bibinfo{author}{\bibfnamefont{C.}~\bibnamefont{Vozzi}},
  \bibinfo{author}{\bibfnamefont{S.}~\bibnamefont{Stagira}}, \bibnamefont{and}
  \bibinfo{author}{\bibfnamefont{M.}~\bibnamefont{Nisoli}},
  \bibinfo{journal}{Phys. Rev. A} \textbf{\bibinfo{volume}{70}},
  \bibinfo{pages}{013411} (\bibinfo{year}{2004}).

\bibitem[{\citenamefont{Pfeifer et~al.}(2007)\citenamefont{Pfeifer, Jullien,
  Abel, Nagel, Gallmann, Neumark, and Leone}}]{Pfeifer2007}
\bibinfo{author}{\bibfnamefont{T.}~\bibnamefont{Pfeifer}},
  \bibinfo{author}{\bibfnamefont{A.}~\bibnamefont{Jullien}},
  \bibinfo{author}{\bibfnamefont{M.~J.} \bibnamefont{Abel}},
  \bibinfo{author}{\bibfnamefont{P.~M.} \bibnamefont{Nagel}},
  \bibinfo{author}{\bibfnamefont{L.}~\bibnamefont{Gallmann}},
  \bibinfo{author}{\bibfnamefont{D.~M.} \bibnamefont{Neumark}},
  \bibnamefont{and} \bibinfo{author}{\bibfnamefont{S.~R.} \bibnamefont{Leone}},
  \bibinfo{journal}{Opt. Express} \textbf{\bibinfo{volume}{15}},
  \bibinfo{pages}{17120} (\bibinfo{year}{2007}).

\bibitem[{\citenamefont{Mansten et~al.}(2009)\citenamefont{Mansten,
  Dahlstr\"om, Mauritsson, Ruchon, L'Huillier, Tate, Gaarde, Eckle, Guandalini,
  Holler et~al.}}]{Mansten2009}
\bibinfo{author}{\bibfnamefont{E.}~\bibnamefont{Mansten}},
  \bibinfo{author}{\bibfnamefont{J.~M.} \bibnamefont{Dahlstr\"om}},
  \bibinfo{author}{\bibfnamefont{J.}~\bibnamefont{Mauritsson}},
  \bibinfo{author}{\bibfnamefont{T.}~\bibnamefont{Ruchon}},
  \bibinfo{author}{\bibfnamefont{A.}~\bibnamefont{L'Huillier}},
  \bibinfo{author}{\bibfnamefont{J.}~\bibnamefont{Tate}},
  \bibinfo{author}{\bibfnamefont{M.~B.} \bibnamefont{Gaarde}},
  \bibinfo{author}{\bibfnamefont{P.}~\bibnamefont{Eckle}},
  \bibinfo{author}{\bibfnamefont{A.}~\bibnamefont{Guandalini}},
  \bibinfo{author}{\bibfnamefont{M.}~\bibnamefont{Holler}},
  \bibnamefont{et~al.}, \bibinfo{journal}{Phys. Rev. Lett.}
  \textbf{\bibinfo{volume}{102}}, \bibinfo{pages}{083002}
  (\bibinfo{year}{2009}).

\bibitem[{\citenamefont{Abel et~al.}(2009)\citenamefont{Abel, Pfeifer, Jullien,
  Nagel, Bell, Neumark, and Leone}}]{ABEL2009}
\bibinfo{author}{\bibfnamefont{M.~J.} \bibnamefont{Abel}},
  \bibinfo{author}{\bibfnamefont{T.}~\bibnamefont{Pfeifer}},
  \bibinfo{author}{\bibfnamefont{A.}~\bibnamefont{Jullien}},
  \bibinfo{author}{\bibfnamefont{P.~M.} \bibnamefont{Nagel}},
  \bibinfo{author}{\bibfnamefont{M.~J.} \bibnamefont{Bell}},
  \bibinfo{author}{\bibfnamefont{D.~M.} \bibnamefont{Neumark}},
  \bibnamefont{and} \bibinfo{author}{\bibfnamefont{S.~R.} \bibnamefont{Leone}},
  \bibinfo{journal}{J. Phys. B} \textbf{\bibinfo{volume}{42}},
  \bibinfo{pages}{075601} (\bibinfo{year}{2009}).

\bibitem[{\citenamefont{Antoine et~al.}(1996)\citenamefont{Antoine, L'Huillier,
  and Lewenstein}}]{Antoine1996}
\bibinfo{author}{\bibfnamefont{P.}~\bibnamefont{Antoine}},
  \bibinfo{author}{\bibfnamefont{A.}~\bibnamefont{L'Huillier}},
  \bibnamefont{and}
  \bibinfo{author}{\bibfnamefont{M.}~\bibnamefont{Lewenstein}},
  \bibinfo{journal}{Phys. Rev. Lett.} \textbf{\bibinfo{volume}{77}},
  \bibinfo{pages}{1234} (\bibinfo{year}{1996}).

\bibitem[{\citenamefont{Nisoli et~al.}(2003)\citenamefont{Nisoli, Sansone,
  Stagira, De~Silvestri, Vozzi, Pascolini, Poletto, Villoresi, and
  Tondello}}]{Nisoli2003}
\bibinfo{author}{\bibfnamefont{M.}~\bibnamefont{Nisoli}},
  \bibinfo{author}{\bibfnamefont{G.}~\bibnamefont{Sansone}},
  \bibinfo{author}{\bibfnamefont{S.}~\bibnamefont{Stagira}},
  \bibinfo{author}{\bibfnamefont{S.}~\bibnamefont{De~Silvestri}},
  \bibinfo{author}{\bibfnamefont{C.}~\bibnamefont{Vozzi}},
  \bibinfo{author}{\bibfnamefont{M.}~\bibnamefont{Pascolini}},
  \bibinfo{author}{\bibfnamefont{L.}~\bibnamefont{Poletto}},
  \bibinfo{author}{\bibfnamefont{P.}~\bibnamefont{Villoresi}},
  \bibnamefont{and} \bibinfo{author}{\bibfnamefont{G.}~\bibnamefont{Tondello}},
  \bibinfo{journal}{Phys. Rev. Lett.} \textbf{\bibinfo{volume}{91}},
  \bibinfo{pages}{213905} (\bibinfo{year}{2003}).

\bibitem[{\citenamefont{Corkum}(1993)}]{Corkum1993}
\bibinfo{author}{\bibfnamefont{P.~B.} \bibnamefont{Corkum}},
  \bibinfo{journal}{Phys. Rev. Lett.} \textbf{\bibinfo{volume}{71}},
  \bibinfo{pages}{1994} (\bibinfo{year}{1993}).

\bibitem[{\citenamefont{Lewenstein et~al.}(1994)\citenamefont{Lewenstein,
  Balcou, Ivanov, L'Huillier, and Corkum}}]{Lewenstein1994}
\bibinfo{author}{\bibfnamefont{M.}~\bibnamefont{Lewenstein}},
  \bibinfo{author}{\bibfnamefont{P.}~\bibnamefont{Balcou}},
  \bibinfo{author}{\bibfnamefont{M.~Y.} \bibnamefont{Ivanov}},
  \bibinfo{author}{\bibfnamefont{A.}~\bibnamefont{L'Huillier}},
  \bibnamefont{and} \bibinfo{author}{\bibfnamefont{P.~B.}
  \bibnamefont{Corkum}}, \bibinfo{journal}{Phys. Rev. A}
  \textbf{\bibinfo{volume}{49}}, \bibinfo{pages}{2117} (\bibinfo{year}{1994}).

\bibitem[{\citenamefont{Balcou et~al.}(1999)\citenamefont{Balcou, Dederichs,
  Gaarde, and L'Huillier}}]{BALCOU1999}
\bibinfo{author}{\bibfnamefont{P.}~\bibnamefont{Balcou}},
  \bibinfo{author}{\bibfnamefont{A.~S.} \bibnamefont{Dederichs}},
  \bibinfo{author}{\bibfnamefont{M.~B.} \bibnamefont{Gaarde}},
  \bibnamefont{and}
  \bibinfo{author}{\bibfnamefont{A.}~\bibnamefont{L'Huillier}},
  \bibinfo{journal}{J. Phys. B} \textbf{\bibinfo{volume}{32}},
  \bibinfo{pages}{2973} (\bibinfo{year}{1999}).

\end{thebibliography}
\end{document}